\documentclass[aps,twocolumn,showpacs]{revtex4}
\usepackage{amssymb,amsbsy,psfig,graphicx}
\begin{document}
%%%%%%%%%%%%%%%%%%%%%%%%%%%
\title{Purity of Gaussian states: measurement schemes
and time--evolution in noisy channels}
%%%%%%%%%%%%%%%%%%%%%%%%%%%
\author{Matteo G. A. Paris$^{1}$}\email{paris@unipv.it}
\author{Fabrizio Illuminati$^{2}$}\email{illuminati@sa.infn.it}
\author{Alessio Serafini$^{2}$}\email{serale@sa.infn.it}
\author{Silvio De Siena$^{2}$}\email{desiena@sa.infn.it}
\affiliation{\mbox{}$^{1}$Quantum Optics \& Information Group, 
INFM UdR di Pavia, Italia \\
\mbox{}$^{2}$Dipartimento di Fisica ``E. R. Caianiello'',
Universit\`a di Salerno, INFM UdR Salerno, INFN Sez. Napoli, G. C. Salerno,
Via S. Allende, 84081 Baronissi (SA), Italia}
%%%%%%%%%%%%%%%%%%%%%%%%%%%
\begin{abstract}
We present a systematic study of the purity for Gaussian
states of single-mode continuous variable systems. We prove the
connection of purity to observable quantities for these states,
and show that the joint measurement of two conjugate
quadratures is necessary and sufficient to determine the purity
at any time.  The statistical reliability and the range of
applicability of the proposed measurement scheme is tested by
means of Monte Carlo simulated experiments.  We then consider the
dynamics of purity in noisy channels.  We derive an evolution
equation for the purity of general Gaussian states both in
thermal and squeezed thermal baths. We show that purity is
maximized at any given time for an initial coherent state
evolving in a thermal bath, or for an initial squeezed state
evolving in a squeezed thermal bath whose asymptotic squeezing is
orthogonal to that of the input state.
\end{abstract}
%%%%%%%%%%%%%%%%%%%%%%%%%%%
\date{April 7, 2003} 
\pacs{03.65.Yz, 42.50.--p, 03.67.Pp, 42.50.Dv}
%%%%%%%%%%%%%%%%%%%%%%%%%%%
\maketitle
%%%%%%%%%%%%%%%%%%%%%%%%%%%
\section{Introduction\label{s:intro}}
Nonclassical features of atomic and radiation systems play a
relevant role in quantum information, communication and high
precision measurements, as well as in many fundamental
experiments to test quantum mechanics \cite{pati,heiss}. In
particular, pure Gaussian states of continuous variable (CV)
systems, such as coherent and squeezed--coherent states, are the
key ingredients of secure optical communication \cite{yuk,yue,gran,ralph}
and Heisenberg limited quantum interferometry \cite{cav,bon,yur,ind,int}.
The characterization of several properties of Gaussian states has
been the subject of intense recent work
\cite{kim,ban,eisert,bin,fil,rad,hol}, stimulated by the seminal
analysis on their entanglement properties \cite{simon,cirac}.
\par
Any attempt to exploit Gaussian states for quantum information
and quantum measurement schemes must however
face the obvious difficulty that pure states are unavoidably
corrupted by the interaction with the environment. Therefore, CV
Gaussian states that are available for experiments are usually
mixed states, and it becomes crucial to establish their degree of
purity (or mixedness) determined by the environmental noise.  In
the present paper, we study the purity of Gaussian states for
single--mode continuous variable systems focusing on two aspects: its
experimental characterization, and its time--evolution in noisy
channels. We first show that the joint detection of two
conjugate quadratures is a necessary and sufficient measurement
to determine the purity of a Gaussian state with reliable
experimental statistics; we then derive an evolution equation for
the purity of Gaussian states in a noisy channel, considering the
instances of a thermal bath and of a squeezed thermal bath, and
determine the evolutions that at any given time maximize the
purity.
\par
Let us refer to $\mu=\hbox{Tr}\left[\varrho^2\right]$ as to the
purity of a generic quantum state $\varrho$; the conjugate
quantity $S_{L} = (1 - \mu)d/(d-1)$, where $d$ is the dimension
of the Hilbert space of the system under investigation, is known
as linear entropy or mixedness.  In general, $\mu$ ranges from
one, which is the value for a pure state, to $\mu=1/d$ for a
completely mixed state. Throughout the paper we will consider CV
systems, {\em i.e.} infinite dimensional Hilbert spaces, and
therefore we will have $0< \mu \leq 1$.  Since $\mu$ is a
nonlinear function of the density matrix, it cannot be connected to 
an observable quantity if we perform repeated measurements on single 
copies of the state.  That is, it cannot
be the expectation value of a single-system self-adjoint operator, 
nor it can be related to a single-system probability distribution 
obtained from a positive operator-valued measure (POVM). On the other 
hand, if collective measurements on two copies of the state are possible, 
then the purity may be measured directly \cite{art,fil2}. For instance,
collective measurements of overlap and fidelity have been experimentally
realized for qubits encoded into polarization states of photons  
\cite{hey}.
\par
In general, purity can be determined by the knowledge of the
quantum state of the system, which in turn can be obtained by
quantum tomography \cite{review}. However, in this case, the
statistics is usually poor, since the measurement of a whole
quorum of observables is needed, unavoidably leading to large
fluctuations \cite{add}.  On the other hand, if we focus our
attention on the class of Gaussian states, it is indeed possible
to find an operative method to experimentally determine $\mu$ .  
In fact, Gaussian states are uniquely defined by their first two
statistical moments, which can be measured by the joint
detection of two conjugate quadratures, say position and
momentum or quadrature phases of the electromagnetic field.
Such a measurement corresponds to an estimate of the Husimi
$Q$--function $Q(\alpha)=\langle\alpha|\varrho |\alpha\rangle$,
$|\alpha\rangle$ being a coherent state of the harmonic
oscillator. We will show that the measurement of the
$Q$--function is the optimal minimal measurement for the purity,
in the sense that it is necessary and sufficient to determine
$\mu$ and requires the minimum number of observables to be
measured.
\par
The joint measurement of two conjugate quadratures is possible 
for a single--mode radiation field as well as for a single atom
\cite{kelly,ste,wil}. Remarkably, for these systems, the class of 
Gaussian states include almost all the states that can be reliably 
produced, and employed in communication or measurement protocols.
\par
Finally, we will show that the previous discussion allows to
unravel the dynamics of purity only in terms of observable
quantities. Indeed, the time--evolution of the purity of an
initial Gaussian state in a noisy channel can be uniquely
expressed as a function of the initial observable parameters of
the input state and of the asymptotic observable parameters of
the environment. This property allows then to determine and
engineer optimal evolutions, i.e.~evolutions that maximize
the purity at any given time. 
\par
The paper is structured as follows. In Section \ref{s:inv} we
show how purity is related to observable quantities for Gaussian
states, and how it can be obtained either from the $Q$--function 
or by three single-quadrature detection. In Section \ref{s:sim} we 
present the results of a systematic numerical analysis, establishing 
the statistical reliability and the range of applicability of the 
method by means of Monte Carlo simulated experimental runs.  
We also show that the $Q$-function based determination of purity
is a more reliable method than single-quadrature
detection.  Section \ref{s:evo} is devoted to
derive and solve an evolution equation for the purity of an
initial Gaussian state in a noisy channel, both for thermal and
squeezed thermal baths. We show that, even though the asymptotic
value of purity is not related to the initial conditions, its
behavior at finite times does depend on the initial squeezing
and thermal excitations, and we determine the
evolutions that maximize the purity at any finite time.
We show in particular that purity is maximized for an
initial coherent state evolving in a thermal bath, or for an initial squeezed
state evolving in a squeezed thermal bath whose asymptotic squeezing is
orthogonal to that of the input state.
Finally, in Section \ref{s:out} we present some concluding remarks.
%%%%%%%%%%%%%%%%%%%%%%%%%%%%%%%%%%%%%%%%%%%%%%%%%%%%%%%%%%%%%%%%%%%%
\section{Purity of Gaussian states\label{s:inv}}
We begin by reviewing some fundamental properties of the Wigner
phase-space representation \cite{barnett} which will be useful
throughout the paper. The Wigner representation of an arbitrary operator
$O$ is defined as follows
\begin{equation}
O(\alpha)=\int_{\mathbb C} \frac{{\rm d}^2\gamma}{\pi^{2}}\:\,{\rm e}^{\bar
\gamma\alpha-\gamma\bar\alpha} \:{\rm Tr}[O\:D(\gamma)]\:,
\end{equation}
where $D(\gamma)=\exp(\gamma  a^{\dag}-\bar\gamma a)$ is the displacement
operator, and ${\rm Tr}[O\:D(\gamma)]$ is usually referred to as the
characteristic function of the operator $O$.  Let $O_{1}$ and $ O_{2}$
be operators that admit regular Wigner representations $O_{1}(\alpha)$
and $O_{2}(\alpha)$. Then the trace $\hbox{Tr}\left[O_{1}
\:O_{2}\right]$ can be computed as an integral over phase space
according to
\begin{equation}
\hbox{Tr}[O_{1}\:O_{2}]=\pi\int_{\mathbb C} {\rm d}^2\alpha\:
O_{1}(\alpha)\:O_{2}(\alpha) {\; .}\label{trace}
\end{equation}
From now on, we will move to the phase--space variables $x$ and 
$p$, corresponding to quadrature phases $\hat{x} = (a + a^{\dag})/\sqrt{2}$ 
and $\hat{p} = i(a^{\dag} - a)/\sqrt{2}$ of the field $a$, 
whose expectation values $\langle \hat{x} \rangle \equiv x$ and 
$\langle \hat{p} \rangle \equiv p$ are related to $\alpha$ by 
$\alpha =( x + ip)/\sqrt{2}$.
\par
The Wigner representation $W(\alpha)$ of the density matrix $\varrho$ of a
quantum state is referred to as the Wigner function of the state.
The class of Gaussian states is defined as the class of states 
with Gaussian Wigner function, namely
\begin{equation}
W(x,p)=\frac{\,{\rm e}^{-\frac{1}{2}X\boldsymbol{\sigma}^{-1}X^{T}}}{\pi\sqrt{{\rm
Det} [\boldsymbol{\sigma}]}}{\:,}\label{wigner}
\end{equation}
where $X$ is the displaced vector $X=\left(x-x_{0},p-p_{0}\right)$ and
$\boldsymbol{\sigma}$ is the covariance matrix
$$\sigma_{ij}=\frac{1}{2}\langle \hat{x}_i \hat{x}_j + 
\hat{x}_j \hat{x}_i \rangle -
\langle \hat{x}_i \rangle \langle \hat{x}_j \rangle \, ,$$ 
where $\hat{x}_1 = \hat{x}, \hat{x}_2 = \hat{p}$.
The density matrix of the most general Gaussian state can be written as
\cite{adam}
\begin{eqnarray}
\varrho= D(\alpha_{0})S(r,\varphi)\nu_{\bar{n}}S^{\dag}(r,\varphi)
D^{\dag}(\alpha_{0}) \, ,
\label{Grho}\;
\end{eqnarray}
where $\alpha_0=(x_0+ip_0)/\sqrt{2}$, and $\nu_{\bar{n}}$ is a thermal state with
average photon number $\bar{n}$:
$$ \nu_{\bar{n}}= \frac{1}{1+\bar{n}}\sum_{k=0}^{\infty}
\left(\frac{\bar{n}}{1+\bar{n}} \right)^k\: |k\rangle\langle k|
\: ,$$
$D(\alpha_{0})$ denotes the displacement operator and $S(r,\varphi)=\exp(
\frac12 r \,{\rm e}^{-i2\varphi} a^{2}-\frac12 r \,{\rm e}^{i2\varphi} a^{\dag 2})$ the
squeezing operator.  A convenient parametrization of Gaussian states can be
achieved replacing the $\sigma_{ij}$'s by $n$, $r$, $\varphi$, which
have a more direct phenomenological interpretation.  By applying the
phase-space representation of squeezing \cite{barnett,walls}, the following
relations are easily derived
\begin{eqnarray}
\sigma_{xx}&=&\frac{2\bar n
+1}{2}\: \Big[\!\cosh(2r)-\sinh(2r)\cos(2\varphi)\Big] \: ,
\nonumber\\
\sigma_{pp}&=&\frac{2\bar n
+1}{2}\:\Big[\!\cosh(2r)+\sinh(2r)\cos(2\varphi)\Big] \: ,
\nonumber \\
\sigma_{xp}&=&\frac{2\bar n +1}{2}\: \sinh(2r)\sin(2\varphi) \: .
\label{vars}
\end{eqnarray}
Exploiting Eq.~(\ref{trace}), one can write
\begin{equation}
\mu \doteq {\rm Tr}[{\varrho}^{2}]= \frac{\pi}{2}\:
\int_{\mathbb R}\int_{\mathbb R}\!{\rm d}x\:{\rm d}p\:
W^{2}(x,p) \: ,
\label{muWig}
\end{equation}
so that, for a Gaussian state
\begin{equation}
\mu=\frac{1}{2 \sqrt{\rm Det [\boldsymbol{\sigma} ]}}
\, = \, \frac{1}{2 \sqrt{\sigma_{xx}\sigma_{pp}-\sigma^{2}_{xp}}} 
{\: .}
\label{dodonov}
\end{equation}
In terms of $\bar{n}$, $r$ and $\varphi$, Eq.~(\ref{dodonov})
can then be recast as \cite{marian1,dodonov}
\begin{equation}
\mu=\frac{1}{2 \bar n +1} \: .
\label{thermo}
\end{equation}
Eq.~(\ref{thermo}) shows that the purity of a generic Gaussian state
depends only on the average number
of thermal photons, as one should expect since displacement and squeezing
are unitary operations. Therefore, the measurement of the purity of a
Gaussian state is equivalent to the measurement of its average number
of thermal photons.
\par
As the last step in connecting $\mu$ to observables we report the 
expression of the $\sigma_{ij}$'s in terms of the $Q$--function $Q(\alpha)$.
This follows from the antinormally ordered expression of the second moments.
We have, for instance
$$
\hat{x}^2=\frac{{ a}^{2}+{ a}^{\dag 2}+2 a a^{\dag}-{\mathbb I}}{2}
\: ,
$$
which, in terms of phase--space variables, 
corresponds to $x^{2}-\frac{1}{2}$.
Therefore, we eventually get
\begin{eqnarray}
\langle \hat{x}^{2} \rangle & = & 
{\rm Tr}[\varrho\:\frac{(a+a^\dag)^{2}}{2}]
 \nonumber \\ 
&=&
\int_{\mathbb R}\int_{\mathbb R}\!{\rm d}x\:{\rm d}p\;Q(x,p)\:(x^{2}-\frac{1}{2})
\nonumber \: ,
\end{eqnarray}
where we have moved from variables $\alpha$ and $\bar\alpha$
to variables $x$ and $p$, previously defined.
In much the same way, we obtain
\begin{eqnarray}
\langle \hat{p}^2 \rangle & = &\int_{\mathbb R}\int_{\mathbb R}
\!{\rm d}x\:{\rm d}p\:Q(x,p)\:(p^{2}-\frac{1}{2}) \, , \\
\frac12 \langle \hat{x}\hat{p} + \hat{p}\hat{x} \rangle & = &
\int_{\mathbb R}\int_{\mathbb R}\!{\rm d}x\:{\rm d}p\:Q(x,p)\:x\:p \: .
\end{eqnarray}
Since first moments are naturally antinormally
ordered, evaluation of first moments of quadratures is easily obtained,
and the $\sigma_{ij}$'s can be eventually computed.
\par
Gaussian states may be effectively characterized as well by single-quadrature 
measurements obtained by balanced homodyne detection \cite{parlik}. 
Thus a question arises whether or not one really needs 
to resort to joint measurement of two conjugate quadratures 
to determine the purity. In particular, 
since Gaussian states are fully characterized by 
the first and second moments,
it suffices to measure the rotated quadrature 
$x_\theta = (a^\dag \,{\rm e}^{i\theta} + a \,{\rm e}^{-i\theta})/\sqrt{2}$ 
for three different values of $\theta$ to have a complete 
characterization of the
state, including the measure of its purity. 
This fact can be proven by reminding that the probability 
distribution $p(x,\theta)$ of a measurement of $x_\theta$ on a
state of the form (\ref{Grho}) is a Gaussian centered in
$x_0=\hbox{Re}[\alpha_0 \,{\rm e}^{-i\theta}]$, with variance 
\begin{equation}
\sigma_{\theta} = \frac{1}{2\mu} \left[\,{\rm e}^{-2r}\cos^2 (\theta -\varphi) 
+\,{\rm e}^{2r}\sin^2 (\theta - \varphi) \right] . 
\end{equation}
By measuring three quadratures we directly obtain the purity $\mu$
by comparison of variances. By choosing
$\theta=0,\pi/2,\pi/4$ 
we have 
\begin{eqnarray}
\mu = \left[4\sigma_{\pi/4}(\sigma_{0} + \sigma_{\pi/2} -
\sigma_{\pi/4}) - (\sigma_{0} - \sigma_{\pi/2})^{2}\right]^{-\frac12} .
\label{tresg}
\end{eqnarray}
In the next Section, we will compare the two different experimental
schemes on the basis of Monte Carlo simulated experiments.
%%%%%%%%%%%%%%%%%%%%%%%%%%%%%%%%%%%%%%%%%%%%%%%%%%%%%%%%%%%%%%%%%%
\section{Monte Carlo simulated experiments}\label{s:sim}
As we have seen, in order to evaluate the $\sigma_{ij}$'s and
then the purity, we need to estimate averages over the
$Q$--function. These estimates can be obtained if one disposes of
data distributed according to the $Q$--function
$Q(x,p)$ itself. Indeed, such a distribution can be
experimentally reconstructed for a single--mode radiation field 
through heterodyne \cite{sha}, eight--port homodyne \cite{ott,leo} 
or six--port homodyne detectors \cite{tri}, and for atoms by 
coupling the atom with two light fields and measuring the 
corresponding phase--shifts \cite{wil}.
\par
%%%%%%%%
\begin{figure}[tb]
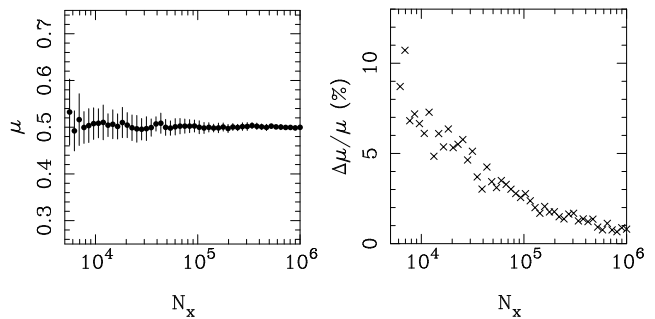

\begin{tabular}{ll}
\psfig{file=varnx_1.ps,width=41mm}&
\psfig{file=varnx_2.ps,width=42mm}\end{tabular}
\caption{Effect of the number of data on the $Q$-function based
determination of purity for Gaussian states: results from Monte Carlo
simulated experiments. On the left graph we plot the determination
of the purity $\mu$ versus the number of data $N_{x}$
for a squeezed thermal state with parameters given by 
$\alpha_0=0$, $\varphi=0$, $r=1.5$, corresponding
to $\sinh^2 1.5\simeq 4.5$ mean squeezed photons, and 
a mean number of thermal photons $\bar{n} = 0.5$.
Black circles are the estimated values of purity based 
on the $Q$--distributed statistics, vertical bars are
the experimental errors (confidence intervals); for a
large number of experimental data the errors
quickly fall well within the black circles of the estimated
values. The theoretical value of purity for all
the simulated experimental runs is $\mu=0.5$. 
On the right graph we report the relative errors
$\Delta\mu/\mu$ versus the number of data for 
the same squeezed thermal state.
\label{f:varnx}}
\end{figure}
%%%%%%%%%
In order to test the effectiveness of the proposed scheme, we have
performed a systematic numerical analysis by means of Monte Carlo
simulated experiments. The simulations are needed to
show the actual independence of the method on the squeezing and
displacing parameters, in compliance with Eq.~(\ref{thermo}). 
Moreover, they provide a crucial test on the
actual possibility of getting reliable (i.e.~with reduced
fluctuations) determinations of $\mu$ in realistic experimental
settings and even for most unfavorable states.
\par
The purity $\mu$ and its dispersion $\Delta\mu$ have been evaluated
from samples of the $Q$--function, 
varying the values of the parameters
of the simulated Gaussian state. Besides
$\bar n$, $r$, $\varphi$, and
$\alpha_{0}$, the experimental determination
of the $Q$--function depends on the number $N_x$ of collected data.
\par
We find that $\mu$ and $\Delta\mu$ are essentially independent on
the complex displacement parameter $\alpha_{0}$ and on the squeezing
angle $\varphi$. On the other hand, $\Delta\mu$ does depend on
$\bar n$ and $r$, decreasing with increasing $\bar{n}$ and increasing 
with increasing $r$.
\par
In Fig.~\ref{f:varnx} we report the determination of purity for a
strongly squeezed thermal state as a function of the number of data.
The error on purity is of the order of a few percent for samples made
of $N_x\simeq 10^5$ data.
%%%%%%%%
\begin{figure}[tb]
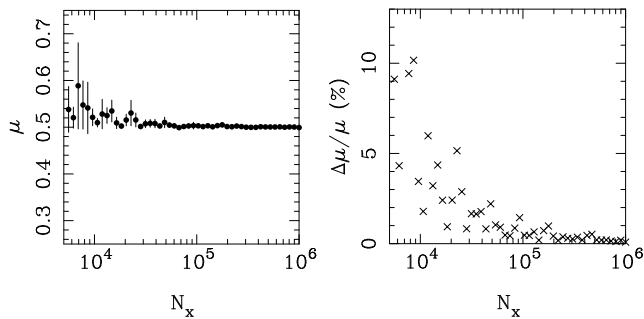

\begin{tabular}{ll}
\psfig{file=trequad_1.ps,width=41mm}&
\psfig{file=trequad_2.ps,width=42mm}\end{tabular}
\caption{Effect of the number of data on the determination of 
purity for Gaussian states by single-quadrature detection: results 
from Monte Carlo simulated experiments of three quadratures $x_0$, 
$x_{\pi/2}$, $x_{\pi/4}$. On the left graph we plot the determination
of the purity $\mu$ versus the number of data $N_{x}$
for the same squeezed thermal state of Fig.~\ref{f:varnx}.
Notice that, in this instance, the total number of data $N_{x}$ 
corresponds to $N_{x}/3$ detections for each quadrature.
Black circles are the estimated values of purity based 
on the balanced homodyne statistics, vertical bars are
the experimental errors (confidence intervals); for a
large number of experimental data the errors
quickly fall well within the black circles of the estimated
values. The theoretical value of purity for all
the simulated experimental runs is $\mu=0.5$. 
On the right graph we report the relative errors
$\Delta\mu/\mu$ versus the number of data for 
the same squeezed thermal state. 
\label{f:trequad}}
\end{figure}
%%%%%%%%%
\par
In order to compare the determination of $\mu$ by the $Q$--function with 
that coming from single--quadrature detection, we have simulated the 
measurement of three quadratures $x_\theta$, 
$\theta=0,\pi/2,\pi/4$ by balanced 
homodyne detection. In Fig.~\ref{f:trequad} we report the estimated purity
[using Eq.~(\ref{tresg})] for the same strongly squeezed thermal state of 
Fig.~\ref{f:varnx} as a function of the number of data. 
Some features are immediately evident. First of all one can see that
the determination is biased: 
in the present case the estimated $\mu$ is always larger than the 
true value, while the opposite case occurs by inverting the phase of 
the squeezing. Therefore the method is very sensitive to the choice
of the phase. Moreover, 
the relative error is not a smooth function of the number of data
i.e.~the method is not statistically reliable as the joint--measurement 
one.
This is again due to the remarkable dependence of the variances on the 
phase of the squeezing, a dependence which is instead smoothed out 
in the measurement of the $Q$--function.
Summing up, for some specific states (as the example considered here) 
single-quadrature detection may be asymptotically even more efficient
than the heterodyne one.
However, in general, the number of data needed for the relative error to be below the 
joint--measurement level is strongly state--dependent. 
We conclude that the measurement of the $Q$--function is statistically more
reliable and thus more suited for a systematic analysis of the purity of 
Gaussian states. 
\par
Let us now go back to the analysis of the $Q$--function determination of purity. 
A smaller number of data is needed to obtain a given precision for
states with smaller squeezing. The effect of the squeezing parameter
on the determination of purity is illustrated in Fig.~\ref{f:varr}, where
we report $\mu$ and $\Delta\mu/\mu$ versus $r$ for Gaussian states with
$\alpha=0$,
$\varphi=0$, and $\bar n=0.5$, and for a number of data
$N_x=3\cdot10^4$. Notice that in Fig.~\ref{f:varr}
the range of $r$ corresponds to a quite large number of mean
squeezed photons $0 \leq \sinh^2 r \lesssim15$.
%%%%%%%%
\begin{figure}[tb]
\begin{tabular}{ll}
\psfig{file=varR_1.ps,width=41mm}&
\psfig{file=varR_2.ps,width=42mm}
\end{tabular}
\caption{Effect of squeezing on the $Q$--function based
determination of purity for Gaussian states: results from Monte Carlo
simulated experiments. On the left graph we plot the determination 
of purity versus the squeezing parameter $r$ for Gaussian states with
the other parameters fixed at $\alpha=0$, $\varphi=0$, and 
$\bar n=0.5$.
Black circles are the determined values of purity based 
on the $Q$--distributed statistics, vertical bars are
the experimental errors (confidence intervals). For small $r$ 
the errors are within the black circles. 
The theoretical value of purity for all
the states is $\mu=0.5$. 
On the right graph we report the relative errors
$\Delta\mu/\mu$ versus the squeezing parameter for the same set of
experiments. The number of data in all simulated experiments
is $N_x=3\cdot10^4$.
\label{f:varr}}
\end{figure}
%%%%%%%%
\begin{figure}[tb]
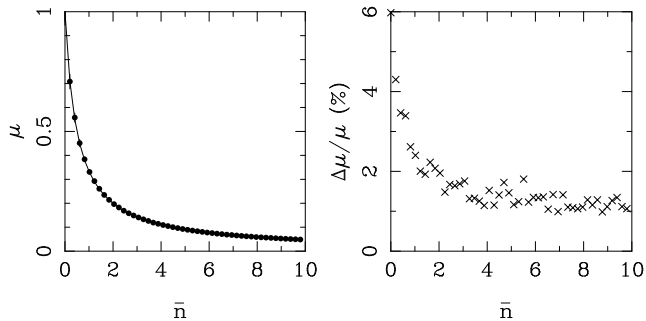

\begin{tabular}{ll}
\psfig{file=varnth_1.ps,width=41mm}&
\psfig{file=varnth_2.ps,width=42mm}
\end{tabular}
\caption{Effect of thermal photons on the $Q$-function based
determination of purity for Gaussian states: results from Monte
Carlo simulated experiments. On the left graph we plot the determination
of purity versus the value of $\bar{n}$ for Gaussian states with
the other parameters fixed at $\alpha=0$, $\varphi=0$, and $r=1.0$.
Black circles are the determined values of purity according to
the $Q$--function statistics, and vertical bars denote
the experimental errors (confidence intervals); the latter are within
the black circles for essentially all values of $\bar{n}$. The solid line
reports the theoretical values of $\mu$. On the right graph we report the
relative errors $\Delta\mu/\mu$ versus $\bar{n}$ for the same
set of experiments. The number of data in all simulated experiments
is $N_x=10^4$.
\label{f:varnth}}
\end{figure}
%%%%%%%%
\par
In the deep quantum regime, i.e.~for small $\bar{n}$,
fluctuations of $\mu$ become more relevant. This is not surprising,
since $\mu$ is a highly nonlinear function of the second--order moments.
However, simulations show that even for highly squeezed (up to $\simeq 15$
mean squeezed photons) and slightly mixed (down to $\bar{n} \simeq 0.1$) 
states, realistic experimental conditions allow a statistically 
reliable determination of $\mu$ that complies with the theoretical 
expectation (\ref{thermo}), up to an error of a few percent.
In Fig.~\ref{f:varnth} we plot the determination of purity for
different squeezed thermal states as a function of 
the average number of thermal photons $\bar{n}$, for samples 
made of $N_x=10^5$ data.
\par
From the above analysis we 
conclude that the joint measurement of two conjugate quadratures
provides a statistically reliable method to
determine the purity of a generic Gaussian state. This is best achieved with
experimental schemes that involve data distributed according to the
Husimi $Q$--function, such as heterodyne and multi--port
homodyne detection schemes.
%%%%%%%%%%%%%%%%%%%%%%%%%%%%%%%%%%%%%%%%%%%%%%%%%%%%%%%%%%%%%%%%%%%%%%
\section{Evolution of purity in a noisy channel\label{s:evo}}
Let us consider the time evolution of an initial, pure or
mixed, generic single--mode Gaussian state in presence of noise 
and damping (and/or pumping) toward a final squeezed thermal state.
If $\Gamma^{-1}$ is the photon lifetime in the noisy channel, 
the evolution of a state is described, in the 
interaction picture, by the following master equation
\begin{eqnarray}
\dot \varrho & = & \frac{\Gamma }{2}N \: L[a^{\dag}]\varrho
+\frac{\Gamma}{2}(N+1)\:L[a]\varrho
\nonumber \\ &-&
\frac{\Gamma}{2}\: \Big( \overline{M}\:D[a]\varrho + M
\:D[a^{\dag}]\varrho \Big)
\label{rhoev}\:,
\end{eqnarray}
where the dot stands
for time--derivative and
the Lindblad superoperators are defined by
\begin{eqnarray}
L[O]\varrho &\equiv&  2 O\varrho O^{\dag} -
O^{\dag} O\varrho -\varrho O^{\dag} O \, , \\
&& \nonumber \\
D[O]\varrho &\equiv & 2 O\varrho O -O O\varrho -\varrho O O
\label{LinSO}\;.
\end{eqnarray}
$M$ is the correlation function of the bath (which is 
usually referred  to as the squeezing of the bath);
it is in general a complex number $M = M_1 + iM_2$,
and $\overline{M}$ denotes its complex conjugate, 
while $N$ is a phenomenological parameter related, 
as we shall see, to
the purity of the asymptotic state. 
Positivity of the density matrix imposes
the important constraint $|M|^{2} \le N(N+1)$.
At thermal equilibrium, i.e.~for $M=0$, $N$
coincides with the average number of thermal photons
in the bath.
The master equation (\ref{rhoev}) can be transformed into a
Fokker--Planck equation for the Wigner function $W(x,p,t)$.  
Using the differential representation of the superoperators 
\cite{walls,paris} in Eq.~(\ref{rhoev}),
the corresponding Fokker-Planck equation reads as follows
\begin{eqnarray}
\dot W (x,p,t) &=& \frac{\Gamma}{2} \Bigg(2+
x \partial_{x}+p \partial_{p}  
+
\frac{2N+1}{2}
\left(\partial_{xx}^{2}+\partial_{pp}^{2}\right) \nonumber \\
&+&
M_{1}\left(\partial_{xx}^{2}-\partial_{pp}^{2}\right)
+
2 M_{2}\partial_{xp}
\Bigg) W(x,p,t) . \label{fokplan}
\end{eqnarray}
For a general single--mode Gaussian state of the
form (\ref{wigner}) one has, in compact notation
\begin{eqnarray}
\dot W  &=& \frac{\Gamma}{2} \Bigg(
2-X\boldsymbol{\sigma}^{-1}{x \choose p}+
\frac{2N+1}{2}\,{\cal S}_{X\boldsymbol{\sigma}}({\mathbb I}) 
\nonumber \\
&+&
M_{1}\,{\cal S}_{X\boldsymbol{\sigma}}({\mathbb A})
+ M_{2}\,{\cal S}_{X\boldsymbol{\sigma}}({\mathbb B})
\Bigg){W} \: ,
\label{wiggo}
\end{eqnarray}
where ${\cal S}_{X\boldsymbol{\sigma}}(\boldsymbol{\gamma})$ 
denotes the seralian (or siralian) operator, a scalar
function of the matrix $\boldsymbol{\gamma}$ given 
by ${\cal S}_{X\boldsymbol
{\sigma}}(\boldsymbol{\gamma})
\equiv X \boldsymbol{\sigma}^{-1}\boldsymbol{\gamma}
\boldsymbol{\sigma}^{-1}X^{T}-
{\rm Tr}[\boldsymbol{\gamma}\boldsymbol{\sigma}^{-1}]$. 
The displaced vector $X$ and the covariance matrix
$\boldsymbol{\sigma}$
have been previously defined, 
whereas $\mathbb {I, A, B}$ form a basis in the space
of $2 \times 2$ real symmetric matrices:
\begin{displaymath}
\mathbb{I} =\left(\begin{array}{cc}
1&0\\
0&1\\
\end{array}\right) \, , \quad
\mathbb{A}=\left(\begin{array}{cc}
1&0\\
0&-1\\
\end{array}\right) \, , \quad
\mathbb{B}=\left(\begin{array}{cc}
0&1\\
1&0\\
\end{array}\right) \, .
\end{displaymath}
For any given real matrix
$\boldsymbol \gamma$ and generic Gaussian states,
the seralian operator shows the remarkable
property:
\begin{equation}
\int_{\mathbb R}{\rm d}x\int_{\mathbb R}{\rm d}p\:{\cal S}_{X\boldsymbol 
\sigma}(\boldsymbol \gamma)\:W(x,p,t)=0 \: .
\end{equation}
It can be easily shown that this property 
assures that the last three terms of Eq.~(\ref{wiggo}) 
[corresponding to diffusion terms in the Fokker--Planck equation (\ref{fokplan})] 
do not enter in
the time--evolution equations for the first statistical moments $x_{0}$ and $p_{0}$.
Such evolution is governed by the drift terms and is described 
by the following equation for the vector $X_{0}\equiv{x_{0} \choose p_{0}}$
\begin{eqnarray}
\dot X_{0}=
\int_{\mathbb R}{\rm d}x\int_{\mathbb R}{\rm d}p\:{x \choose p}
\frac{\dot W}{2} &=& -\frac{\Gamma}{2}X_{0}
\: .
\end{eqnarray}
First moments are damped through the noisy channel: this 
effect should be expected since it is the mathematical evidence
of the absorption of the state's coherent photons.\par 
The evolution of the covariance matrix of the state can be described by
monitoring different sets of variables. A good choice of variables
is given by the $\sigma_{ij}$'s, in terms of which the evolution
equations decouple.
The relations between the variables $\sigma_{ij}$ 
and the variables $\mu$, $r$, and $\varphi$ are given in
Eqns.~(\ref{vars}) and (\ref{dodonov}). Here we recall
some further relations that will be useful in the following:
\begin{equation}
{\rm Det}[\boldsymbol{\sigma}]=
\sigma_{xx}\sigma_{pp}-\sigma_{xp}^{2}=\frac{(2\bar{n}
+1)^{2}}{4}=\frac{1}{4\mu^{2}} {\rm \; ,}\label{det}
\end{equation}
\begin{equation}
{\rm
Tr}[\boldsymbol{\sigma}]=\sigma_{xx}+\sigma_{pp}
=(2\bar{n}+1)\cosh(2r)=
\frac{\cosh(2r)}{\mu} {\rm \; ,} \label{tr}
\end{equation}
\begin{equation}
\sigma_{pp}-\sigma_{xx}=\frac{\sinh(2r)\cos(2\varphi)}{\mu}
\: . \label{dif}
\end{equation}
As we have seen, in the Wigner phase--space picture the expectation
values can be computed as phase--space integrals. The 
first--order evolution equation for the covariance matrix
$\boldsymbol{\sigma}$
is thus obtained by straightforward integration, and reads:
\begin{eqnarray}
\dot{\boldsymbol{\sigma}}&=&\Gamma\left(
\boldsymbol{\sigma}_{\infty}-\boldsymbol{\sigma}\right)\,{\rm ,}
\label{mateq}\\
&&\nonumber\\
{\rm with}\quad\boldsymbol{\sigma}_{\infty}&\equiv&\left(\begin{array}{cc}
\frac{(2N+1)+2M_{1}}{2}&M_{2}\\
&\\
M_{2}&\frac{(2N+1)-2M_{1}}{2}\end{array}\right)
\, . \label{asycov} 
\end{eqnarray}
The matrix $\boldsymbol{\sigma}_{\infty}$, 
determined by the bath parameters alone, turns out to be 
the asymptotic covariance matrix. In fact, integration 
of Eq.~(\ref{mateq}) yields
\begin{equation}
\boldsymbol{\sigma}(t)=\boldsymbol{\sigma}_{\infty}
\left(1-{\rm e}^{-\Gamma t}\right)
+\boldsymbol{\sigma}(0)\,{\rm e}^{-\Gamma t}.
\label{solution}
\end{equation}
Eq.~(\ref{solution}) shows a simple example of 
a Gaussian completely positive map \cite{eisplen}. 
The Gaussian character of the evoluted Wigner 
function can be proven, {\em a posteriori}, by 
verifying that a function 
of the form (\ref{wigner}), with covariance matrix given by 
Eq.~(\ref{solution}), 
indeed solves Eq.~(\ref{fokplan}). In order to be a 
{\it bona fide}
covariance matrix, $\boldsymbol{\sigma}(t)$ must satisfy
the usual 
condition encoding the $\hat{x}-\hat{p}$ 
uncertainty relations \cite{eisplen, simsud}
\begin{equation}
\boldsymbol{\sigma}(t)+\frac{i}{2}{\bf J}\ge 0{\, ,}
\quad{\rm with}\quad {\bf J}=\left(\begin{array}{cc}
0&1\\-1&0\end{array}\right) \, . 
\end{equation}
It is promptly seen that such a condition is satisfied
at any time by the convex combination 
giving $\boldsymbol{\sigma}(t)$ in Eq.~(\ref{solution}) iff 
$\boldsymbol{\sigma}_{\infty}$ is a legitimate
covariance matrix. 
This last requirement is assured by the necessary constraint
$N(N+1)\ge|M|^{2}$ that guarantees positivity of the density
matrix.  
\par
\begin{figure}[b!]
\includegraphics[width=8cm]{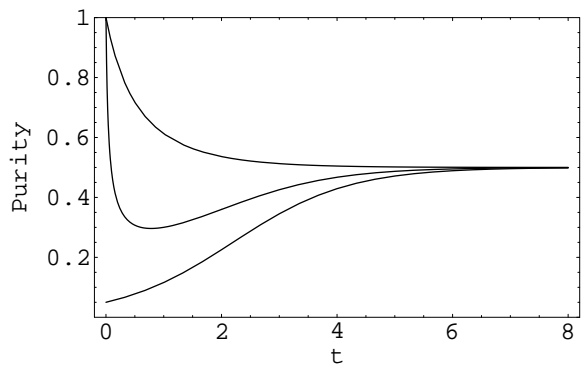}
\caption{\small The purity $\mu$ for various Gaussian states 
evolving in a channel with $N=0.5$, $M=0$, as a function of time.
Time is dimensionless and measured in units of $\Gamma^{-1}$. 
The upper curve refers to an initial pure coherent state ($r_0=0$,
$\mu_0=1$), the central curve to an initial pure squeezed vacuum 
($r_{0}=1.5$, $\mu_0=1$), and the lower curve to an initial
thermal state with $r_{0}=0$ and $\mu_{0}=0.05$, i.e.~
${\bar{n}}_{0}=9.5$.}
\label{purtempo}
\end{figure}

By introducing 
$$\mu_{\infty}\doteq
\Big[(2N+1)^{2}-4|M|^{2}\Big]^{-1/2} \: ,$$ and exploiting
Eqns.~(\ref{det}--\ref{dif}) we can eventually
express $\mu$, $r$ and $\varphi$ as functions of time
\newpage
\begin{widetext}
\begin{eqnarray}
\mu(t)&=&\mu_{0}\bigg[\frac{\mu_{0}^{2}}{\mu_{\infty}^{2}}\left(1-
{\rm e}^{-\Gamma t}\right)^{2} \, + \,
{\rm e}^{-2\Gamma t} \nonumber \\
&& \nonumber \\
&+&2\mu_{0}\Big(\frac{\sqrt{1+4\mu_{\infty}^{2}
|M|^{2}}\cosh(2r_{0})}{\mu_{\infty}}
+2\sinh(2r_{0})\big(M_{1}\cos(2\varphi_{0})
-M_{2}\sin(2\varphi_{0})\big)
\Big)
\left(1-{\rm e}^{-\Gamma t}\right){\rm e}^{-\Gamma t}
\bigg]^{-1/2}{\rm
\, ,} \label{purtot}
\end{eqnarray}
\begin{equation}
\cosh[2r(t)]=
\mu(t)\left(\frac{\sqrt{1+4\mu_{\infty}^{2}|M|^{2}}\left(1-
{\rm e}^{-\Gamma t}\right)}{\mu_{\infty}}+
{\rm e}^{-\Gamma t}\frac{\cosh(2r_{0})}{\mu_{0}}\right){\rm
\, ,} \label{squiztot}
\end{equation}
\begin{equation}
\tan[2\varphi(t)]=\frac{M_{2}2\mu_{0}\left(1-{\rm e}^{-\Gamma t}\right)
+\sinh(2r_{0})\sin(2\varphi_{0}){\rm e}^{-\Gamma t}}
{-M_{1}2\mu_{0}\left(1-{\rm e}^{-\Gamma t}\right)+
\sinh(2r_{0})\cos(2\varphi_{0}){\rm e}^{-\Gamma t}}{\rm
\: ,} \label{phitot}
\end{equation}
where $\mu_{0}$, $r_{0}$ and $\varphi_{0}$ are,
respectively, the initial purity and the initial
squeezing parameters.
\par
Let us first consider the case $M=0$,
for which the initial state is
damped toward a thermal state with mean photon number $N$
\cite{barnett,marian}. In this case, see Eq.~(\ref{phitot}),
$\varphi$ is constant in time and does not enter in the expression
of $\mu$.
The corresponding solutions for $\mu(t)$ and $r(t)$ read 
then as follows
\begin{eqnarray}
\mu(t)&=& \mu_{0} \left[\frac{\mu_{0}^{2}}{\mu_{\infty}^{2}}
\left(1-{\rm e}^{-\Gamma t}\right)^{2}+2\frac{\mu_{0}}{\mu_{\infty}}{\rm
e}^{-\Gamma t}
\left(1-{\rm e}^{-\Gamma t}\right)\cosh(2r_{0})+{\rm e}^{-2
\Gamma t}\right]^{-1/2} \: ,
\label{purezza} \\
&& \nonumber \\
\cosh[2r(t)]&=&
\mu(t)\left(\frac{1-{\rm e}^{-\Gamma t}}{\mu_{\infty}}+
{\rm e}^{-\Gamma t}\frac{\cosh(2r_{0})}{\mu_{0}}\right){\rm
\, .} \label{squiz}
\end{eqnarray}
\end{widetext}
The quantities $\mu(t)$ and $r(t)$ in Eqns.~(\ref{purezza}) 
and (\ref{squiz}) solve the following system of coupled 
equations
\begin{eqnarray}
\dot{\mu}&=&\Gamma \left(\mu-\frac{\mu^{2}\cosh(2r)}{\mu_{\infty}}
\right) \: , \nonumber\\
&& \nonumber \\
\dot r&=&-\frac{\Gamma}{2}\frac{\mu}{\mu_{\infty}}\sinh(2r)
{\rm \; ,} 
\label{system}
\end{eqnarray}
which, in turn, can be directly found working 
out the basic evolution equation $\dot \mu=2{\rm Tr}[\dot
\varrho\: \varrho]$ as a phase--space integral and 
exploiting Eqns.~(\ref{det}--\ref{dif}).  
It is easy to see that, as $t\rightarrow \infty$,
$\mu(t)\rightarrow\mu_{\infty}= (2N+1)^{-1}$ and $r(t)\rightarrow 0$, as
one expects, since the channel damps (pumps)
the initial state to a thermal state with
mean photon number $N$.
Therefore, the only constant solution of Eq.~(\ref{system}) is
$\mu=\mu_{\infty}$, $r=0$, i.e.~only initial non--squeezed states 
are left unchanged by the evolution in the noisy channel.
%%%%%%
\begin{figure}[tb]
\psfig{file=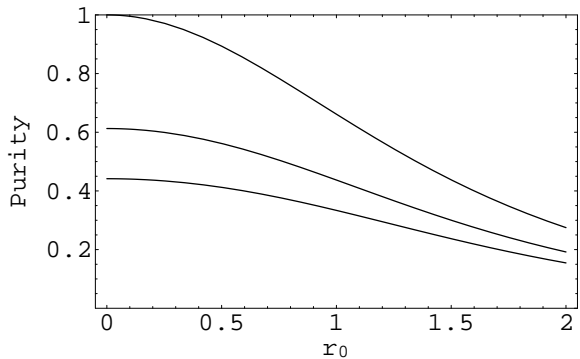,width=8cm}
\caption{Plot of the purity $\mu$ 
for an initially pure Gaussian state ($\mu_{0}=1$)
in different non--squeezed ($M=0$) noisy channels, 
evaluated at time $t=\Gamma^{-1}$, as a
function of the initial squeezing parameter $r_{0}$.
From top to bottom, the value of the mean thermal photon
number $N$ that characterizes the
different channels is $N=0$, $N=0.5$, and
$N=1$ respectively.\label{pursquiz}}
\end{figure}
%%%%%
In fact Eq.~(\ref{purezza}) shows that $\mu(t)$ is a decreasing function
of $r_{0}$: in a non--squeezed channel ($M=0$), a squeezed state 
decoheres more rapidly than a non--squeezed one (see Figs.~\ref{pursquiz}
and \ref{purtempo}).
Let us consider, for instance, an initially pure state in a channel with 
$N=1$ (so that $\mu_{\infty}=\frac{1}{3}$);
after a time $t=\Gamma^{-1}$, the ratio of the
purity of a state with $r_{0}=1.5$ to the purity of a state with $r_{0}=0$ is
53.7\%. This dependence could therefore be relevant for practical
purposes.
The optimal evolution for the purity,
obtained letting $r=0$ in Eq.~(\ref{purezza}), reads
\begin{equation}
\mu(t)=\frac{\mu_{0}\:
\mu_{\infty}}{\mu_{0}+{\rm e}^{-\Gamma t}(\mu_{\infty}-\mu_{0})}
\: . \label{optpur}
\end{equation}
\par
Obviously, $\mu(t)$ is not necessarily a decreasing
function of time: if $\mu_{0} < \mu_{\infty}$
then the initial state will undergo a certain amount
of purification, asymptotically reaching the value
$\mu_{\infty}$ which characterizes the channel,
as shown in Fig.~\ref{purtempo}.
In addition, $\mu(t)$ is not a
monotonic function for any choice of the initial conditions. 
Letting $\dot{\mu} = 0$ in Eq.~(\ref{system}), and exploiting 
Eqns.~(\ref{purezza}) and (\ref{squiz}), one
finds the following condition for the appearance of a zero of $\dot \mu$ at
finite positive times: $\cosh(2r_{0}) > {\rm max}
(\frac{\mu_{0}}{\mu_{\infty}}, \frac{\mu_{\infty}}{\mu_{0}})$.
If this condition is satisfied, 
then $\mu(t)$ shows a local extremum, in fact a minimum
since, differentiating the first of Eqns.~(\ref{system}) 
and letting $\dot \mu=0$, one obtains 
$\ddot \mu > 0$. This behavior is shown in Fig.~\ref{purtempo}.
%%%%%
%%%%%
\par
Let us now treat the more
general instance $M \neq 0$ of a squeezed thermal bath. 
Recalling the definition of $\mu_{\infty}$
and exploiting Eqns.~(\ref{purtot}--\ref{phitot}), one easily finds
the asymptotic values of the physical parameters 
$\mu$, $r$ and $\varphi$
\begin{eqnarray}
\mu_{\infty}&=&\frac{1}{\sqrt{(2N+1)^{2}-4|M|^{2}}} 
\: , \label{purasi} \\
&& \nonumber \\
\cosh(2r_{\infty})&=&\sqrt{1+4\mu_{\infty}^{2}|M|^{2}}
\: , \label{squizasi} \\
&& \nonumber \\
\tan(2\varphi_{\infty})&=&-\frac{M_{2}}{M_{1}}
\: . \label{phiasi}
\end{eqnarray}
These values characterize the squeezed channel.
Eq.~(\ref{purasi}) shows that,
if $M\neq0$, then $N$ is not simply the mean thermal photon 
number $\bar{n}$ of the asymptotic state. One has:
$$ N = \frac{\sqrt{\left( 2\bar{n} + 1\right)^{2} +
4|M|^{2}} - 1}{2} \: . $$
\par
In order to understand the dynamics of
purity when $M \neq 0$, it is convenient to
write again the expression (\ref{purtot})
for $\mu(t)$, using Eqns.~(\ref{squizasi}) and (\ref{phiasi}) to
switch from the complex parameter $M=M_{1}+iM_{2}$ to 
the asymptotic values of the squeezing parameters
$r_{\infty}$ and $\varphi_{\infty}$; one obtains
\begin{widetext}
\begin{eqnarray}
\mu(t)&=&\mu_{0}\bigg[\frac{\mu_{0}^{2}}{\mu_{\infty}^{2}}\left(1-
{\rm e}^{-\Gamma t}\right)^{2} \, + \,
{\rm e}^{-2\Gamma t} \nonumber \\
&+&2\frac{\mu_{0}}{\mu_{\infty}}\Big(\cosh(2r_{\infty})\cosh(2r_{0})
+\sinh(2r_{\infty})\sinh(2r_{0})\big(\cos(2\varphi_{\infty}-2\varphi_{0})
\big)
\Big)
\left(1-{\rm e}^{-\Gamma t}\right){\rm e}^{-\Gamma t}
\bigg]^{-1/2}{\, .}
\label{mubathsqueez}
\end{eqnarray}
We see from Eq.~(\ref{mubathsqueez}) that
$\mu(t)$ is a monotonically decreasing function of the
factor $\cos(2\varphi_{\infty}-2\varphi_{0})$, which
gives the only dependence on the initial phase $\varphi_{0}$
of the squeezing.
Thus, for any given $\varphi_{\infty}$ characterizing the
squeezing of the bath, 
$\varphi_{0}=\varphi_{\infty}+\frac\pi 2$ is
the most favorable value of the initial angle of squeezing,
i.e.~the one which allows the maximum purity at a given time.
For such a choice, $\mu(t)$ reduces to
\begin{equation}
\mu(t)=\mu_{0}\bigg[\frac{\mu_{0}^{2}}{\mu_{\infty}^{2}}\left(1-
{\rm e}^{-\Gamma t}\right)^{2}+
{\rm e}^{-2\Gamma t}
+2\frac{\mu_{0}}{\mu_{\infty}}\cosh(2r_{\infty}-2r_{0})
\left(1-{\rm e}^{-\Gamma t}\right){\rm e}^{-\Gamma t}
\bigg]^{-1/2}{\: \rm .}
\end{equation}
\end{widetext}
This is a decreasing function of the factor $\cosh(2r_{\infty}-2r_{0})$,
so that the maximum value of the purity at a given time is achieved
for the choice $r_{0}=r_{\infty}$, and the evolution of
the purity of a squeezed state in a squeezed channel is
identical to the evolution of the purity of a non--squeezed 
state in a non--squeezed channel expressed by Eq.~(\ref{optpur})
and illustrated in Fig.~\ref{purtempo}.\par
In conclusion,
for the most general instance of a channel characterized
by arbitrary $\mu_{\infty}$, $r_{\infty}$,
$\varphi_{\infty}$ and $\Gamma$, the initial Gaussian state
for which purity is best preserved in time
must have a squeezing parameter $r_{0}=r_{\infty}$ and
a squeezing angle $\varphi_{0}=\varphi_{\infty}+\frac\pi 2$,
i.e.~it must be antisqueezed (orthogonally squeezed)
with respect to the bath.
The net effect for the evolution of the purity is that
the two orthogonal squeezings of the initial state and of
the bath cancel each other exactly, thus reproducing the optimal
purity evolution of an initial non--squeezed coherent state 
in a non--squeezed thermal bath.
%%%%%%%%%%%%%%%%%%%%%%%%%%%%%%%%%%%%%%%%%%%%%%%%%%%%%%%%%%%%%%%%%%%%%%
\section{Conclusions\label{s:out}}
We have shown that the purity of Gaussian states 
for continuous variable systems
can be operationally determined by the joint measurement of two 
conjugate quadratures. In order to perform such a measurement, 
the minimal, necessary and
sufficient requirement is that the measurement apparatus 
records data distributed according to the Husimi 
quasi--probability function. We have then verified
by Monte Carlo simulated experiments the statistical 
reliability of the associated
measurement schemes, thus proving the possibility of an experimentally
realizable characterization of the purity of Gaussian states.
We have compared as well the scheme based on the $Q$--function
with the one based on single--quadrature detection, and showed
that the former provides a more reliable statistics.
Moreover, we have derived an evolution equation for the purity of
Gaussian states in noisy channels, both in the case of a thermal and 
of a squeezed thermal bath.
Our analysis shows that the purity is
maximized at any given time for an initial coherent state
evolving in a thermal bath, or for an initial squeezed state
evolving in a squeezed thermal bath whose squeezing is
orthogonal to that of the input state.
We have focused our attention on the purity of 
single--mode Gaussian states.
The time--evolution of the purity for specific initial non
Gaussian states of great physical relevance can be studied,
as well as the extension to Gaussian states of multimode
systems, both pure and mixed. 
These topics are currently being explored and
will be the subject of forthcoming work.  
%%%%%%%%%%%%%%%%%%%%%%%%%%%%%%%%%%%%%%%%%%%%%%%%%%%%%%%%%%%%%%%%%%%%%%
\acknowledgments{The work of MGAP has been sponsored by INFM through the 
project PRA-2002-CLON. FI, AS and SDS thank 
INFM and INFN for financial support. SDS thanks the ESF COSLAB program.
FI thanks the ESF BEC2000+ program.}
%%%%%%%%%%%%%%%%%%%%%%%%%%%%%%%%%%%%%%%%%%%%%%%%%%%%%%%%%%%%%%%%%%%%%%

%%%%%%%%%%%%%%%%%%%%%%%%%%%%%%%%%%%%%%%%%%%%%%%%%%%%%%%%%%%%%%%%%%%%%%

\end{document}